 \documentclass[acmsmall,screen]{acmart}

\AtBeginDocument{%
  \providecommand\BibTeX{{%
    \normalfont B\kern-0.5em{\scshape i\kern-0.25em b}\kern-0.8em\TeX}}}

\setcopyright{rightsretained}
\acmJournal{CSUR}
\acmYear{2024} \acmVolume{1} \acmNumber{1} \acmArticle{1} \acmMonth{4}\acmDOI{10.1145/3659945}

\acmBooktitle{ACM Computing Surveys} 
\acmPrice{15.00}
\acmISBN{978-1-4503-XXXX-X/18/06}

\usepackage{enumitem}

\usepackage{soul}
\usepackage{xcolor}

\usepackage{booktabs}
\usepackage{multirow}
\usepackage{graphicx}

\usepackage{fontawesome}   

\usepackage{xspace}

\usepackage{pifont}
\let\oldding\ding
\renewcommand{\ding}[2][1]{\scalebox{#1}{\oldding{#2}}}

\usepackage{arydshln}
\usepackage{colortbl}
\makeatletter
\def\adl@drawiv#1#2#3{%
 \hskip.5\tabcolsep
 \xleaders#3{#2.5\@tempdimb #1{1}#2.5\@tempdimb}%
 #2\z@ plus1fil minus1fil\relax
 \hskip.5\tabcolsep}
\newcommand{\cdashlinelr}[1]{%
 \noalign{\vskip\aboverulesep
  \global\let\@dashdrawstore\adl@draw
  \global\let\adl@draw\adl@drawiv}
 \cdashline{#1}
 \noalign{\global\let\adl@draw\@dashdrawstore
  \vskip\belowrulesep}}
\makeatother

\definecolor{palered}{HTML}{f4e1c2}
\definecolor{paleorange}{HTML}{e08c1c}
\definecolor{paleyellow}{HTML}{fdffb6}
\definecolor{palegreen}{HTML}{b7c88b}
\definecolor{darkpalegreen}{HTML}{627337}  
\definecolor{palecyan}{HTML}{9bf6ff}
\definecolor{paleblue}{HTML}{708cb4}
\definecolor{darkpaleblue}{HTML}{415a7c}   
\definecolor{palepurple}{HTML}{bdb2ff}
\definecolor{palepink}{HTML}{ffc6ff}
\definecolor{darkpalepink}{HTML}{880088}  
\definecolor{palewhite}{HTML}{fffffc}

\begin{document}

\title[A Systematic Analysis of XR Deceptive Design]{Deceived by Immersion: A Systematic Analysis of Deceptive Design in Extended Reality}  %


\author{Hilda Hadan}
\email{hhadan@uwaterloo.ca}
\orcid{https://orcid.org/0000-0002-5911-1405}
\affiliation{
    \institution{Stratford School of Interaction Design and Business, University of Waterloo}
    \city{Waterloo}
    \country{Canada}
}

\author{Lydia Choong}
\email{ljmchoong@uwaterloo.ca}
\orcid{https://orcid.org/0009-0006-1279-9085}
\affiliation{
    \institution{Cheriton School of Computer Science, University of Waterloo}
    \city{Waterloo}
    \country{Canada}
}

\author{Leah Zhang-Kennedy}
\email{lzhangke@uwaterloo.ca}
\orcid{https://orcid.org/0000-0002-0756-0022}
\affiliation{
    \institution{Stratford School of Interaction Design and Business, University of Waterloo}
    \city{Waterloo}
    \country{Canada}
}

\author{Lennart E. Nacke}
\email{lennart.nacke@acm.org}
\orcid{https://orcid.org/0000-0003-4290-8829}
\affiliation{
    \institution{Stratford School of Interaction Design and Business, University of Waterloo}
    \city{Waterloo}
    \country{Canada}
}

\renewcommand{\shortauthors}{Hadan, et al.}

\begin{abstract}
The well-established deceptive design literature has focused on conventional user interfaces. With the rise of extended reality (XR), understanding deceptive design's unique manifestations in this immersive domain is crucial. However, existing research lacks a full, cross-disciplinary analysis that analyzes how XR technologies enable new forms of deceptive design. Our study reviews the literature on deceptive design in XR environments. We use thematic synthesis to identify key themes. We found that XR's immersive capabilities and extensive data collection enable subtle and powerful manipulation strategies. We identified eight themes outlining these strategies and discussed existing countermeasures. Our findings show the unique risks of deceptive design in XR, highlighting implications for researchers, designers, and policymakers. We propose future research directions that explore unintentional deceptive design, data-driven manipulation solutions, user education, and the link between ethical design and policy regulations.

\footnotetext{We would like to thank graduate researchers Sabrina Alicia Sgandurra and Derrick Wang for their insightful feedback on the manuscript and their expertise in resolving technical issues during our paper formatting. We also thank the reviewers for their constructive criticism that ensured a smooth and polished final publication of our research. The research was supported by the Natural Sciences and Engineering Research Council of Canada (NSERC) Discovery Grant (\#RGPIN-2022-03353 and \#RGPIN-2023-03705), the Social Sciences and Humanities Research Council of Canada (SSHRC) Insight Grant (\#435-2022-0476), the Canada Foundation for Innovation (CFI) JELF Grant (\#41844). Any opinions, findings, and conclusions or recommendations expressed in this material are those of the author(s) and do not necessarily reflect the views of the NSERC, SSHRC, the CFI, nor the University of Waterloo. 

Authors' Address: H. Hadan, L. Choong, L. Zhang-Kennedy; emails: \{hhadan, ljmchoong, lzhangkennedy\}@uwaterloo.ca, L. E. Nacke; email: lennart.nacke@acm.org.
}
\end{abstract}





\maketitle

\section{Introduction}
\label{sec:introduction}

Extended Reality (XR), which comprises Virtual Reality (VR), Augmented Reality (AR), and Mixed Reality (MR) technologies, has generated substantial research and industry interest---especially in the games industry---since 2012~\cite{Oculus2012Kickstarter}. XR is currently experiencing rapid growth~\cite{statista2022XR}. The literature has highlighted the potential of XR to enhance gaming and socialization~\cite{torstensson2020wizard,thomas2012survey}, arts and design~\cite{schlembach2021forced}, e-commerce advertisements~\cite{mhaidli2021identifying}, and education~\cite{nijholt2012trends}. The rise of XR technology has prompted discussions on deceptive design\footnote{We follow the ACM Diversity and Inclusion Council's guideline for inclusive language and adopt the term ``deceptive design'' instead of ``dark patterns'' in our study. See: \url{https://www.acm.org/diversity-inclusion/words-matter}} (also known as ``dark pattern'')---the user interface design that researchers deem manipulative~\cite{Brignull2022deceptive,mathur2021makes}---from experts in engineering~\cite{lee2021adcube,su2022perception}, security and privacy~\cite{buck2022security, giaretta2022security}, cognitive science~\cite{franklin2022virtual}, and humanities and social science~\cite{schlembach2021forced}. 

The use of XR technology enables new deception opportunities that are not present in other digital environments. For example, e-commerce companies can induce artificial emotions and target vulnerable users to influence their purchasing decisions~\cite{mhaidli2021identifying}. Recently, researchers from different disciplines have begun to study XR deceptive designs and propose corresponding solutions. However, a high-level research overview of deceptive design strategies in XR is lacking. More specifically, there is limited research that addresses how XR and deceptive design tactics influence each other, and what new harm XR deceptive designs bring to users. While scholarly work focusing on XR deceptive design is still in its infancy, we believe that it is critical to seed discussions and provide analytical clarity on the deceptive design strategies identified thus far. Since XR technologies are not yet widely adopted by consumers, researchers, designers, and policymakers can benefit from early constructive feedback and design recommendations aimed at reducing the potential harms of XR technologies.

With its inclusion in XR technologies, concerns about deceptive design have extended to personal security~\cite{su2022perception, lee2021adcube}, safety~\cite{cummings2022all}, social~\cite{ramirez2021what, schlembach2021forced}, ethical~\cite{ramirez2021what,schlembach2021forced}, and political~\cite{schlembach2021forced} aspects during and after engagement with XR. XR uses various technologies that interact with multiple human senses (e.g., hearing, vision, touch) to deliver an immersive experience. Thus, deceptive design could be even more problematic in XR than in non-XR environments because users are more deeply engaged and, therefore, more prone to manipulations~\cite{franklin2022virtual}. 

Although extensive research on deceptive design has already elucidated the harms in non-XR environments~\cite{mathur2021makes}, its use in XR technology is still not well understood. XR manufacturers are already introducing advertisements (e.g., Meta announced the experiment of in-headset VR ads~\cite{Meta2021Testing}), and other deceptive design tactics will soon follow. Thus, reviewing current knowledge on XR deceptive design and providing a guideline for future research is an urgent necessity. Delaying research until XR deceptive design is a more developed field would expose users to risks that have not yet been studied. In this work, we synthesize XR deceptive design strategies identified in previous work to derive important insights for researching and designing solutions for deceptive design and develop new ways for the safe and ethical use of XR technologies.

In our literature analysis, we began by asking how previous works defined deceptive design in their research. Despite the growing focus on XR, there is little literature on deceptive design in this context. This gap indicates an urgent need. We must understand how deceptive design manifests in XR, because its unique traits affect user manipulation. Thus, to build a strong foundation, our \textbf{RQ1} asks, \textit{how has the existing literature defined deceptive design in the context of XR?}
Deceptive design found on websites, games, and mobile apps often relies on interface design elements (e.g., a countdown timer)~\cite{Brignull2023book,gray2018dark}. Research indicates that XR's immersive capabilities~\cite{XRSI2020Definition}, including multi-sensory feedback~\cite{buck2022security, mhaidli2021identifying}, have the potential to modify users' choice architecture~\cite{selinger2010competence}. Therefore, \textbf{RQ2} asks, \textit{How can XR amplify the effects of deceptive design?} XR has unique potential for manipulation with sensory feedback. It is vital to investigate if new deceptive designs, not relying on visual elements, might emerge in this domain. Thus, \textbf{RQ3:} asks, \textit{What deceptive design strategies can be present in XR?} With the possibility of both amplified and novel deceptive design tactics in XR, we must understand the potential risks this poses to users. Therefore, \textbf{RQ4} asks, \textit{what risks can XR deceptive design pose to users?}

To answer our research questions, our research methodology involves a \textit{systematic analysis} of the literature using \textit{thematic synthesis}. We considered 187 candidate articles from a search in four bibliography databases from various research disciplines. After following the rigorous and systematic process of ~\citet{page2021prisma}, we screened the abstracts and full papers, resulting in the elimination of 31 duplicates and 143 articles that solely discussed deceptive design \textit{or} XR. Our final sample comprised 13 articles that discussed the application of deceptive design in XR technologies. 

We make the following \textbf{contributions}: \textit{First}, we provide an analysis of how deceptive design is defined within XR research. We focus on distinguishing nudging from persuasion. This promotes conceptual clarity and shows where future deceptive design research should focus. Particularly, we emphasize unintentional user decisions. \textit{Second}, we show how XR's immersive qualities can be uniquely exploited for deception, providing concrete examples of how manipulation may differ from other platforms. \textit{Third}, present a categorization of deceptive design strategies found in XR through our thematic analysis. \textit{Fourth}, we analyze the interplay between deceptive design and existing XR risks, revealing how they mutually exacerbate potential user harms. Finally and \textit{fifth}, we synthesize prevention techniques from XR studies and provide actionable recommendations for XR designers, policymakers, and educators to mitigate deceptive design risks. Through this work, we propose future research directions to study XR deceptive design from unintentional design decisions and the difference between manipulative (e.g., tricking) and benevolent (e.g., nudging) design strategies, understand how XR poses risks to users, develop more transparent XR data use practices, and create educational strategies on deceptive design practices in XR.

\section{Theoretical Background \& Motivation} 
\label{sec:background}
Deceptive design\footnote{\label{note2} See footnote 1.} (also known as ``dark patterns'') gained attention in 2010 when \citet{Brignull2022deceptive} first introduced the concept. Deceptive design generally describes design patterns that distort or impair users' decision-making ability, making them engage in undesired behaviors or make choices they would not otherwise make~\cite{mathur2021makes,Brignull2023book,EUDSA2022}. Researchers have focused on its applications in various fields. including websites, mobile apps, games, and gamification~\cite{zagal2013dark,dillon2020digital,lewis2014irresistible}. However, it's investigation in XR environments had been limited. \textbf{XR} encompasses Virtual Reality (VR), Augmented Reality (AR), and Mixed Reality (MR)~\cite{mhaidli2021identifying}. \textbf{VR} lets users interact with virtual objects by providing visual and auditory feedback in a fully immersive virtual environment~\cite{ramirez2021what, su2022perception, buck2022security, XRSI2020Definition}. By creating a virtual world that replaces reality, VR blocks users' perception of the real world~\cite {mhaidli2021identifying}. Unlike VR, \textbf{AR} overlays virtual content onto the physical world to enhance the environment, rather than completely obscuring it~\cite{lebeck2018towards, buck2022security}. Compared to VR and AR, the conception of \textbf{MR} is varied~\cite{speicher2019mixed}. In this work, we adopt the most widely used MR explanation from ~\citet{speicher2019mixed} and \citet{XRSI2020Definition}, that is, MR offers hybrid reality, which denotes an environment in which both physical and virtual settings coexist and interact in real-time. VR/AR/MR technologies such as headsets and controllers (i.e., Oculus Rift\footnote{Meta Oculus Rift S.~\url{https://www.oculus.com/rift-s/}} and Microsoft HoloLens\footnote{Microsoft HoloLens.~\url{https://www.microsoft.com/en-us/hololens}}), smart glasses (e.g., Magic Leap\footnote{Magic Leap.~\url{https://www.magicleap.com/magic-leap-2}}), handheld devices (e.g., smartphone), and virtual projectors are all considered XR~\cite{XRSI2020Definition}. The technology creates the potential for new forms of manipulation that have not been investigated in detail in the literature.
   
\subsection{Harms of Deceptive Design in Non-XR and XR Environments}

Research has widely documented the concerns raised by deceptive design in various digital contexts~\cite{lewis2014irresistible, zagal2013dark, dillon2020digital, gray2021consent}. Deceptive design undermines user \textit{autonomy} when making decisions~\cite{mathur2021makes,susser2019online} and potentially causes them to suffer \textit{financial loss} such as paying for things they did not mean to~\cite{Brignull2022deceptive,zagal2013dark}, or limit their ability to make informed buying decisions~\cite{Brignull2022deceptive,bongard2021manipulated}. Users can also experience \textit{privacy risks} from bad defaults, hard-to-access privacy-respecting options, or designs that take advantage of their fear and drive them away from privacy-respecting options~\cite{mathur2021makes,Brignull2022deceptive,Forbrukerradet2018deceived,gray2023mapping}. The data collected from devices can also become an exploitable source to detriment users~\cite{Greenberg2014proxemic}. Moreover, some deceptive designs create \textit{cognitive burden} such as requiring the users to spend extra time, energy, and attention to obstruct them from selecting the desired choices (i.e., hard-to-cancel subscriptions~\cite{Brignull2022deceptive} and cookie consent with hard-to-find deny options~\cite{soe2020circumvention,gray2021consent}). Users' susceptibility to deceptive design is influenced by factors such as the frequency of its occurrence, its misleading behavior and UI appearance, as well as the perceived trustworthiness and level of frustration of the users~\cite{m2020towards}. Despite users being aware of the impact of deceptive design, their uncertainty about the actual harm discourages them from taking self-protective actions~\cite{bongard2021manipulated}. In addition, users' dependency on services can lead to a resigned attitude, making it hard for them to avoid deceptive designs~\cite{maier2019dark}.

XR's sophisticated tracking capabilities create unprecedented risks for amplifying the harms of deceptive design~\cite{buck2022security,mhaidli2021identifying}.
XR sensors track minute user movements, generating rich visual, audio, and haptic feedback~\cite{buck2022security}. These data empower inferences about users' physical and mental states~\cite{o2016convergence,miller2020personal}, habitual patterns~\cite{miller2020personal,pfeuffer2019behavioural, maloney2021social}, and even cognitive, emotional, and personal vulnerabilities~\cite{buck2022security, mhaidli2021identifying, o2016convergence,o2023privacy}. Users often reveal biometric and demographic data to achieve optimal XR functionality~\cite{maloney2020anonymity}. Bad actors can weaponize this sensitive information for deanonymization and targeted manipulation~\cite{miller2020personal, pfeuffer2019behavioural, gugenheimer2022novel}.
Manufacturers and third-party companies could potentially exploit these deceptive design tactics for profit~\cite{egliston2021examining}, altering users' behaviors, emotions, and decision-making processes~\cite{mhaidli2021identifying,buck2022security,o2016convergence}. Despite these dangers, our understanding of how deceptive design manifests within this unique context remains limited. Additionally, users often lack awareness of XR data collection practices and their potential misuse~\cite{o2023privacy,hadan2024privacy}, hindering their ability to take self-protective measures~\cite{o2016convergence}. Given XR's rapid adoption in diverse spheres of life~\cite{cummings2022all}, research must urgently examine XR deceptive design beyond e-commerce and advertising. We must critically examine its implications for social interactions, political manipulation, and other potentially harmful applications.

\subsection{Expanding Deceptive Design Taxonomies and Characteristics}
\label{subsec:deceptive_design}

The taxonomies on deceptive design are large and growing over the years. Building upon Harry Brignull's 2010 effort, \citet{zagal2013dark} outlined four major categories of deceptive game design in 20,000 mobile games. In 2014, \citet{Greenberg2014proxemic} analyzed deceptive design that exploits proxemic interactions and detriments users. In addition, in their 2018 study, \citet{gray2018dark} further enriched the Brignull's 2010 classifications into a taxonomy with five major categories. In 2019, \citet{mathur2019dark} conducted a large-scale analysis of shopping websites and identified seven categories of deceptive design. In the same year, \citet{fitton2019F2P} and \citet{karlsen2019exploited} investigated deceptive design in mobile apps and web-based games, drawing inspiration from~\citet{zagal2013dark}. Based on~\citet{gray2018dark}'s taxonomy, \citet{geronimo2020UI} studied user perception of deceptive design in mobile games and apps. \citet{gray2021consent} discussed the deceptive design in cookie consent banners. In 2023, \citet{Brignull2023book} expanded \citet{mathur2019dark}'s classification into eight deceptive strategies and seven deceptive design types. Building upon this, \citet{mildner2023defending} and \citet{mildner2023engaging} examined deceptive design presented in social media platforms. Furthermore, \citet{king20233d} investigated players' perception of deceptive design on 3D interfaces, and \citet{roffarello2023defining}'s literature review identified eleven types of deceptive design that are capable of capturing user attention.

Combining the diverse deceptive design taxonomies identified in previous research, \citet{mathur2021makes} summarized six deceptive design attributes: (1) \textit{Asymmetric} designs that impose unequal burdens on the choices available to the users (e.g., choices that benefit the service are feature prominently), (2) \textit{Covert} designs that push users toward certain options using mechanisms that users cannot recognize, (3) \textit{Deceptive} designs that induce false beliefs in users through misleading statements and intentional omissions, (4) \textit{Information hiding} designs that obscure or delay the presentation of information that is necessary for decision-making to users, (5) \textit{Restrictive} designs that reduce or eliminate the choices available to users, and (6) \textit{Disparate treatment} designs that treat a particular group of users different from others (e.g., provide additional resources after payment~\cite{zagal2013dark}). All these attributes modify users' \textit{Choice Architectures} (i.e., the decision making-space~\cite{thaler2013choice}) and effectively manipulate users~\cite{Adeyoju2022privacy,thaler2013choice}. \citet{gray2023towards} further synthesized the deceptive design taxonomies in academic literature and regulatory frameworks (e.g.,~\cite{EUDSA2022,Brignull2023book,mathur2019dark,gray2018dark,luguri2021shining}), and developed a domain-agnostic ontology with six high-level deceptive patterns: nagging, obstruction, sneaking, interface interference, forced action, and social engineering~\cite{gray2023towards}.

With the existing taxonomies of deceptive design in mind, our research aims to identify the deceptive mechanisms and characteristics that are prevalent in XR environments, so that similarities and differences to other platforms can be identified and studied.

\section{Methodology}
\label{sec:method}

\subsection{Systematic Analysis of the Literature}
\label{subsec:sys_lit_rev}

Our systematic analysis methodology for database searching, screening, and extraction was based on PRISMA~\cite{page2021prisma} due to its clear guidelines that better facilitate literature review and meta-analysis processes. We started with $187$ initial unique publication records and arrived at a final sample of $13$ papers. To answer our RQs, we then synthesized data by Thematic Synthesis~\cite{thomas2008methods}. This section provides the details of our systematic analysis. We present our thematic synthesis approach in~\autoref{subsec:thematic_syn}. 

\begin{figure}[!t]
  \includegraphics[width=\textwidth]{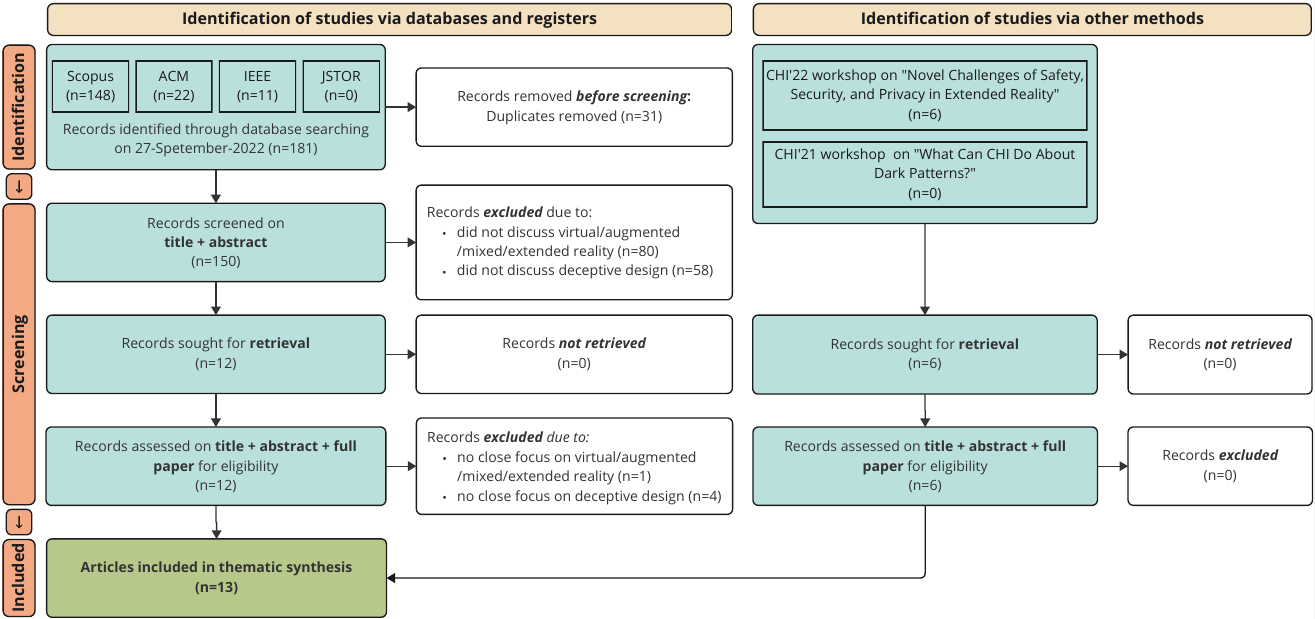}
  \caption{This PRISMA flow diagram~\cite{page2021prisma} presents all the phases of our systematic analysis of the literature, from the identification of articles to the final articles we included.}
  \Description{This PRISMA flow diagram presents all the phases of our systematic analysis of the literature, from the identification of articles to the final articles we included. We reviewed 187 articles from four bibliography databases and two workshop. After following the rigorous and systematic process of~\citet{page2021prisma}, we eliminated all but 13 articles that closely focused on XR deceptive design.}
  \label{fig:PRISMA}
\end{figure}

\subsubsection{Databases, Search Queries, and Duplicate Removal}
\label{subsubsec:database_and_search}

We began with the development of the search protocol, where we defined the search query and databases. To ensure the inclusiveness of our search, we targeted four bibliographic databases: Scopus\footnote{\url{https://www.scopus.com/}}, the ACM Guide to Computing Literature\footnote{\url{https://libraries.acm.org/digital-library/acm-guide-to-computing-literature}}, IEEE Xplore\footnote{\url{https://ieeexplore.ieee.org/Xplore}}, and JSTOR\footnote{\url{https://www.jstor.org/}}. Both the ACM Guide to Computing Literature and IEEE Xplore offer a strong focus on technology- and engineering-related publications, Scopus has multi-discipline publications, and JSTOR provides access to humanities and social sciences journals. Given that XR is a cluster of technological innovations, and deceptive design takes advantage of human weaknesses, we decided to include all four databases in our initial search. The search results from all four databases together offered a good balance in depth and breadth for our review. 

In addition, we defined a set of search terms using keywords that frequently appeared in deceptive design definitions (as described in \autoref{subsec:deceptive_design}), as well as terms frequently used to describe XR technology. We conducted multiple iterations of searches with these keywords to test and refine the combinations of search terms and to ensure that the research outcomes fit our scope. \autoref{tab:search_query} presents the final search query for individual databases.   


Our search was based on ``Abstracts'' instead of ``Full Text'', given that Scopus and JSTOR do not allow ``Full Text'' searches. We intended to keep our search query as consistent as possible. Thus, the ``Abstract'' search was applied to all four databases. Further, our search terms excluded \textit{nouns} such as ``manipulation'' and ``deception'' because these terms were frequently used to describe manipulating factors in controlled experiments or skills in surgical procedures (e.g., laparoscopic manipulation skill), and to describe brain activities in neuroscience (e.g., brain activity in deception and truth-telling). Moreover, acronyms of XR technologies and terms such as ``trick'', ``steer'', ``mislead'' and ``subvert preferences'' were excluded, since these were frequently used in other disciplines to represent irrelevant topics. Our final search queries include terms that describe XR technology: ``Virtual Reality,'' ``Augmented Reality,'' ``Mixed Reality,'' ``Extended Reality;'' and terms that describe deceptive design: ``dark pattern,'' ``deceptive design,'' ``deceive,'' ``manipulative,'' ``abusive.'' We included the detailed search query for each database in \autoref{tab:search_query}. 

\begin{table}[!h]
\centering
\resizebox{\textwidth}{!}{%
\begin{tabular}{@{}l|l@{}}
\toprule
\textbf{Database} & \textbf{Search Query} \\ \midrule
\begin{tabular}[c]{@{}l@{}}The ACM Guide to \\ Computing Literature*\end{tabular} & \begin{tabular}[c]{@{}l@{}}Abstract: [(``dark patterns'' OR ``dark pattern'' OR ``deceive'' OR ``deceptive'' OR ``manipulative'' OR ``abusive'') \\AND (``extended reality'' OR ``virtual reality'' OR ``augmented reality'' OR ``mixed reality'')]\end{tabular} \\ \midrule
Scopus & \begin{tabular}[c]{@{}l@{}}[ABS** (``dark pattern'') OR ABS (``dark patterns'') OR ABS (``deceive'') OR ABS (``deceptive'') OR ABS (``manipulative'')\\ OR ABS (``abusive'')] AND [ABS (``extended reality'') OR ABS(``virtual reality'') OR ABS (``augmented reality'')\\ OR ABS (``mixed reality'')] AND (LIMIT-TO (DOCTYPE , ``cp'') OR LIMIT-TO (DOCTYPE , ``ar'')\\ OR LIMIT-TO (DOCTYPE , ``re'') OR LIMIT-TO (DOCTYPE , ``ch'')] AND (LIMIT-TO (LANGUAGE , ``English'')]\end{tabular} \\ \midrule
IEEE Xplore* & \begin{tabular}[c]{@{}l@{}}
([(``Abstract'':``dark pattern'') OR (``Abstract'':``dark patterns'') OR (``Abstract'':``deceive'') OR \\
(``Abstract'':``deceptive'') OR (``Abstract'':``manipulative'') OR (``Abstract'':``abusive'')] AND 
[(``Abstract'':``extended reality'') \\OR (``Abstract'':``virtual reality'') OR (``Abstract'':``augmented reality'') OR
(``Abstract'':``mixed reality'')] 

\end{tabular} \\ \midrule
JSTOR* & \begin{tabular}[c]{@{}l@{}}[(``dark pattern'') OR (``dark patterns'') OR (``deceive'') OR (``deceptive'') OR (``manipulative'') OR (``abusive'')] \\ AND [(``extended reality'') OR (``virtual reality'') OR (``augmented reality'') OR (``mixed reality'')] AND la:(eng OR en)\end{tabular} \\ \bottomrule
\multicolumn{2}{l}{\begin{tabular}[c]{@{}l@{}} \textit{Note.} *To ensure the consistency of research queries across the four databases, the corresponding filters for research article, extended abstract, \\ short paper, book chapter, conference paper, book review, or journal paper, and English language were applied through interface features. \\ **``ABS'' is the search syntax for abstract search in Scopus library.
\end{tabular}} \\ 
\end{tabular}%
}
\vspace{1mm}
\caption{This table lists the final search queries in the syntax of each database. The filters we set up for all databases were kept as consistent as possible. The search was conducted twice over each database in September 23 and 27, 2022.}
\label{tab:search_query}
\end{table}

Lastly, we gathered a total of 187 publications comprised of 181 articles from bibliographic databases and 6 workshop papers. We downloaded the titles and abstracts of the $187$ publications in a spreadsheet. We manually inspected each and excluded $31$ duplicates with identical titles and DOIs. In the end, we arrived at $n=156$ unique records for the next phase.

\subsubsection{Screening and Eligibility Scoping}
\label{subsubsec:screening_and_criteria}

With three coders, we first screened the $156$ publications' titles and abstracts based on the following \textit{exclusion criteria}. Then, each publication was screened by two coders. In case of disagreement, the disagreed publications ($4$ publications) were passed on to the third coder who acted as a tie-breaker.

\begin{itemize}
    \item The paper is not about VR/AR/MR technologies or XR technologies as a whole,
    \item The paper is not about deceptive design or any design that manipulates users,
    \item The paper is not about the application of deceptive design in XR technologies,
    \item The paper is not a ``Research Article'', ``Extended Abstract'', ``Short Paper'', ``Book Chapter'', ``Conference Paper'', ``Book Review'', nor ``Journal Paper'',
    \item The paper is not in English,
    \item The full text of the paper is not retrievable. 
\end{itemize}

A total of $18$ publications were retained after the screening process. We then conducted a full-text screening of each, following the same exclusion criteria. Two coders screened the publications individually, and the third coder acted as a tie-breaker. We excluded $5$ publications that only mentioned XR deceptive design as an example in the abstract but did not closely discuss the details in the paper. We retained $n=13$ records after the screening process. The final sample includes $7$ records from the four bibliographic databases and $6$ records from the CHI workshops. We then conducted a thematic synthesis on these $13$ publications. 

\subsection{Thematic Synthesis Approach}
\label{subsec:thematic_syn}

\begin{figure}[!ht]
\centering
  \includegraphics[width=\textwidth]{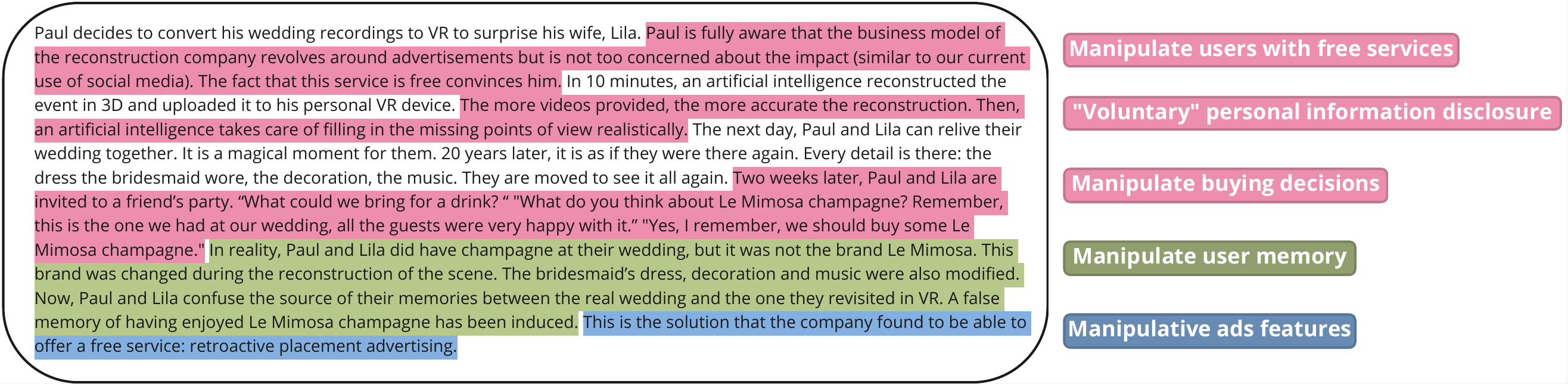}
  \caption{Example of our line-by-line coding process on a snippet from~\citet{bonnail2022exploring}. From left to right: the original snippet, and the respective thematic coding. The colour of the codes correspond to the respective themes in~\autoref{tab:paper-themes}.}
  \Description{Example of our line-by-line coding process on a snippet from~\citet{bonnail2022exploring}. From left to right: the original snippet, and the respective thematic coding. The colour of the codes correspond to the respective themes in~\autoref{tab:paper-themes}.}
  \label{fig:dovetail_example}
\end{figure}

We used thematic synthesis, a rigorous method for identifying and analyzing themes within qualitative data, based on thematic analysis~\cite{clarke2015thematic} for identifying and developing ``themes'' from literature~\cite{thomas2008methods}. Our process involved four stages: (1) data extraction, (2) coding text, (3) developing descriptive themes, and (4) developing analytical themes~\cite{thomas2008methods}.
\begin{enumerate}
    \item The first step of our thematic synthesis process was data extraction. We downloaded the PDFs and extracted the content of $13$ publications into text files using Adobe Acrobat PDF reader. These $13$ text files were then uploaded to Dovetail\footnote{\url{Dovetailapp.com}} for our thematic synthesis. 
    \item Two coders independently conducted line-by-line coding of all publications using an inductive approach~\cite{clarke2015thematic}. During the first round of coding, the two coders independently coded the first $2$ publications. A third coder resolved disagreements and facilitated consensus building. By repeating this process, the two coders each coded the remaining 11 publications.
    \item Initial codes were discussed in weekly meetings, leading to a collaboratively refined codebook. After four weeks, we finalized a codebook with $114$ codes. See ~\autoref{fig:dovetail_example} for an illustrative example of our line-by-line coding process.
    \item We grouped codes into descriptive themes that closely summarized the content of the $13$ publications. Through affinity mapping, discussion, and iterative refinement, we developed eight analytical themes that directly addressed our research questions. \autoref{tab:paper-themes} maps these final themes to each publication. 
\end{enumerate}
We present our results in the following sections.

\section{Results}
\label{sec:results}
\begin{table}[!t]
\centering
\resizebox{\textwidth}{!}{%
\begin{tabular}{@{}lllllllllllllllll@{}}
\toprule
\multicolumn{2}{l|}{} & \multicolumn{4}{c|}{Database (count)} & \multicolumn{3}{c|}{Publication Type (count)} & \multicolumn{3}{c|}{Metrics (average)} & \multicolumn{5}{c}{Countribution Type (count)}\\ \hline
Year & \multicolumn{1}{l|}{Total} & ACM* & IEEE* & Scopus & \multicolumn{1}{l|}{Other**} & Conference & Journal & \multicolumn{1}{l|}{Workshop} & Pages & Authors & \multicolumn{1}{l|}{Citations} & Empirical & Artifact & Theory & Literature & Argument\\
2012 & \multicolumn{1}{l|}{1} & 1 & 0 & 1 & \multicolumn{1}{l|}{0} & 0 & 1 & \multicolumn{1}{l|}{0} & 16 & 13 & \multicolumn{1}{l|}{7} & 0 & 1 & 0&0 & 0 \\
2018 & \multicolumn{1}{l|}{1} & 0 & 1 & 1 & \multicolumn{1}{l|}{0} & 1 & 0 & \multicolumn{1}{l|}{0} & 17 & 4 & \multicolumn{1}{l|}{85} & 1 & 0&0&0&0 \\
2020 & \multicolumn{1}{l|}{1} & 1 & 0 & 1 & \multicolumn{1}{l|}{0} & 1 & 0 & \multicolumn{1}{l|}{0} & 15 & 4 & \multicolumn{1}{l|}{1} & 0 & 1&0&0&0 \\
2021 & \multicolumn{1}{l|}{4} & 2 & 0 & 4 & \multicolumn{1}{l|}{0} & 2 & 2 & \multicolumn{1}{l|}{0} & 17.5 & 3.3 & \multicolumn{1}{l|}{7.5} & 0& 2& 1& 1 & 0 \\
2022 & \multicolumn{1}{l|}{6} & 0 & 0 & 0 & \multicolumn{1}{l|}{6} & 1 & 0 & \multicolumn{1}{l|}{5} & 3.8 & 2.7 & \multicolumn{1}{l|}{0.5} & 0& 1 & 0& 0& 5 \\ \hline
 & \multicolumn{1}{l|}{13} & 4 & 1 & 7 & \multicolumn{1}{l|}{6} & 5 & 3 & \multicolumn{1}{l|}{5} & 10.8 & 3.8 & \multicolumn{1}{l|}{9.7} & 1 & 5 & 1 & 1 & 5 \\ \hline
\multicolumn{17}{l}{\begin{tabular}[c]{@{}l@{}}\textit{Note.} Publications for the year 2022 are only included until May 2022. Aggregated values are counts for the Database and Publication Type columns, and averages for the \\Metrics columns. Citation  numbers were retrieved from Google Scholar on November 25, 2022. \\ *Selected publications from IEEE Xplore and ACM Digital Library also appeared in Scopus. We removed the duplicates but labeled these publications as from multiple sources. \\ **``Other'' includes publications from the 1st Workshop on Novel Challenges of Safety, Security, and Privacy in Extended Reality.\end{tabular}}
\end{tabular}
}
\caption{Overview of the selected publications by year.}
\label{tab:overview}
\end{table}
The first paper relating to XR deceptive design was published in November 2012, after which there was a five-year gap before the second publication in 2018. Since 2020, there has been an increase in the number of publications, reaching a peak of 6 in May 2022. This trend reflects the growth in the development and use of XR technology and increasing research attention on deceptive design since 2010. Our search in Scopus found 4 ACM Digital Library and 1 IEEE Xplore publications among the 13 publications. In total, five publications contributed an artifact, including functional systems, prototypes, or hypothetical scenarios of deceptive XR interfaces. Five workshop short abstracts contributed an argument that discusses the significance of deceptive design problems in XR. Only a few publications made empirical, theoretical, or literature review contributions. Most researchers published at conferences or workshops (5 out of 13) because technology innovation moves quickly. We present a summary of the databases, types of publications, paper metrics, and types of contributions in \autoref{tab:overview}.

\subsection{RQ1: Overview of Deceptive Design Definitions in XR Research}
\label{subsec:overview}
Overall, existing studies on deceptive design in XR largely leverage strategies and patterns identified from previous studies in the web, PC, and mobile app contexts, but extend them by exploring novel implementations in the immersive nature of XR environments. Similar to non-XR deceptive design definitions, the literature on deceptive design in XR often distinguished ``deception'' from ``nudging'' and ``persuasion.'' For example, in their paper on manipulative XR advertising, \citet{mhaidli2021identifying} distinguished persuasion and manipulation by user interests. Specifically, \citet{mhaidli2021identifying} defined ``persuasion'' as users making product purchases by examining and debating advertisers' information. However, ``deception'' was defined as advertisers manipulating consumers into doing things they do not want to or otherwise would not do. Deception and nudging changed user decisions. However, ``nudging'' affects the decision-making context to influence people's decisions without changing their preferences. It lets people ``go their own way.''~\cite[p.~530]{ramirez2021what}. Deception misleads and manipulates users.   

Previous literature has emphasized the role of designer intent in orchestrating deceptive design~\cite{Brignull2022deceptive, gray2018dark}. Three papers (i.e., \citet{krauss2022exploring,cummings2022all,mhaidli2021identifying}) described deceptive design based on definitions from previous studies. For example, \citet{krauss2022exploring} adopted the early definition of deceptive design from \citet{Brignull2022deceptive} used for websites and apps. Similarly, \citet{cummings2022all} adopted the deceptive design definition from \citet{gray2018dark}, which is broader and UX-practitioner-focused. We also observed several papers discussing the benevolent use of deceptive strategies in XR, such as a game to teach children about online risks by incorporating deceptive game components that get participants to share personal information~\citet{torstensson2020wizard}. In VR sports training, virtual characters were constructed with deceptive body actions (e.g., fake body movements to hide the final running direction). This increased players' capacity to recognize or replicate these moves against actual opponents. In the same study, \citet{torstensson2020wizard} discussed deceitful behaviours and dialogues of virtual characters for social training, which mimics real-world negotiation tactics with uncooperative opponents. Overall, we did not identify any publications focused on redefining deceptive design based on use cases from XR environments. In recent deceptive design scholarship, researchers have concluded that user manipulation may not always be intentional~\cite{gray2023mapping,Brignull2023book,EUDSA2022}.  However, only 2 papers briefly hypothesized the possible occurrence of deceptive design from non-manipulative design decisions~\cite{krauss2022exploring,ramirez2021what}. In our work, we focus on analyzing the deceptive design characteristics instead of the intention of the XR designer.

Our systematic analysis of this literature derived eight main themes and 15 subthemes summarized in~\autoref{tab:paper-themes} that describe unique deceptive design strategies in XR. In reporting the results below, the high-level themes in the headings correspond to the themes (T1–T8) summarized in Table~\ref{tab:paper-themes}. The sub-themes related to T2, T3, T5, and T8 are labeled in-line (e.g., T2.1, T2.2).


\begin{table}[!t]
\centering
\resizebox{\textwidth}{!}{%
\begin{tabular}{@{}llllllllllllllll@{}}
\toprule
\multicolumn{3}{l}{\textbf{Reference}} & \rotatebox{90}{\citet{nijholt2012trends}} & \rotatebox{90}{\citet{lebeck2018towards}} & \rotatebox{90}{\citet{torstensson2020wizard}} & \rotatebox{90}{\citet{schlembach2021forced}} & \rotatebox{90}{\citet{lee2021adcube}} & \rotatebox{90}{\citet{mhaidli2021identifying}} & \rotatebox{90}{\citet{ramirez2021what}} & \rotatebox{90}{\citet{krauss2022exploring}} & \rotatebox{90}{\citet{cummings2022all}} & \rotatebox{90}{\citet{buck2022security}} & \rotatebox{90}{\citet{bonnail2022exploring}} & \rotatebox{90}{\citet{su2022perception}} & \rotatebox{90}{\citet{franklin2022virtual}} \\ \midrule
\textbf{Year}& \multicolumn{2}{l}{\textbf{Theme (T)}} & 2012 & 2018 & 2020 & 2021 & 2021 & 2021 & 2021 & 2022 & 2022 & 2022 & 2022 & 2022 & 2022 \\ \midrule
\rowcolor{paleblue!10}\textbf{XR Effects} & \multicolumn{2}{l}{\textcolor{darkpaleblue}{\textbf{T1:}} Adverse effects of user experience} & \faCheck & \faCheck & - & \faCheck & \faCheck & \faCheck & \faCheck & \faCheck & - & \faCheck & \faCheck & - & \faCheck \\
\rowcolor{paleblue!10} & \multicolumn{2}{l}{\textcolor{darkpaleblue}{\textbf{T2:}} Exacerbated user manipulation:} &   &  &   &   &  &   &   &   &   &   &   &   &  \\
\rowcolor{paleblue!10} &  & \textcolor{darkpaleblue}{\textbf{T2.1:}}~\emph{the illusion of objectivity in XR experience} & - & - & - & - & - & - & \faCheck & - & - & - & - & - & - \\
\rowcolor{paleblue!10} &  & \textcolor{darkpaleblue}{\textbf{T2.2:}}~\emph{obscuring reality, disguising risks} & - & \faCheck & - & - & - & - & - & - & - & \faCheck & - & \faCheck & - \\
\rowcolor{paleblue!10} &  & \textcolor{darkpaleblue}{\textbf{T2.3:}}~\emph{data fuels privacy risks and manipulation} & - & \faCheck & - & - & \faCheck & \faCheck & - & - & \faCheck & \faCheck & - & - & - \\
\rowcolor{paleblue!10} &  & \textcolor{darkpaleblue}{\textbf{T2.4:}}~\emph{undesired access, undesired data use} & - & \faCheck & - & - & - & - & - & - & \faCheck & \faCheck & - & - & \faCheck \\
\rowcolor{paleblue!10}  &  & \textcolor{darkpaleblue}{\textbf{T2.5:}}~\emph{insecurity worsen user manipulation and privacy concerns} & - & - & - & - & \faCheck & - & - & - & - & \faCheck & - & \faCheck & - \\
\rowcolor{paleblue!10} &  & \textcolor{darkpaleblue}{\textbf{T2.6:}}~\emph{persistent exposure to manipulation} & - & - & - & - & - & \faCheck & - & - & \faCheck & - & - & - & \\
 \cdashlinelr{2-16} 
\rowcolor{palegreen!10} \textbf{Strategies} & \multicolumn{2}{l}{\textcolor{darkpalegreen}{\textbf{T3:}} Psychological manipulation:} & & & & & & & & & & & & & \\
\rowcolor{palegreen!10} &  & \textcolor{darkpalegreen}{\textbf{T3.1:}}~\emph{false memory implantation} & - & - & - & \faCheck & - & - & - & - & - & - & \faCheck & - & - \\
\rowcolor{palegreen!10} &  & \textcolor{darkpalegreen}{\textbf{T3.2:}}~\emph{artificial prosthetic memory and empathy-based manipulation} & - & - & - & \faCheck & - & \faCheck & \faCheck & - & - & - & - & - & - \\
\rowcolor{palegreen!10} &  & \textcolor{darkpalegreen}{\textbf{T3.3:}}~\emph{hyperpersonalization} & - & - & - & - & - & \faCheck & - & - & - & \faCheck & - & - & - \\
\rowcolor{palegreen!10} & \multicolumn{2}{l}{\textcolor{darkpalegreen}{\textbf{T4:}} Reality distortion} & - & \faCheck & - & - & - & \faCheck & \faCheck & - & \faCheck & - & \faCheck & - & \faCheck \\
\rowcolor{palegreen!10} & \multicolumn{2}{l}{\textcolor{darkpalegreen}{\textbf{T5:}} User perception tricking:} & & & & & & & & & & & & & \\
\rowcolor{palegreen!10} &  & \textcolor{darkpalegreen}{\textbf{T5.1:}}~\emph{blurry boundary between virtual and reality} & - & - & - & - & - & \faCheck & - & - & - & \faCheck & - & - & - \\
\rowcolor{palegreen!10} &  & \textcolor{darkpalegreen}{\textbf{T5.2:}}~\emph{perception hacking} & - & - & - & - & \faCheck & - & - & - & - & - & - & \faCheck & - \\ \cdashlinelr{2-16} 
\rowcolor{palepink!10} \textbf{Risks} & \multicolumn{2}{l}{\textcolor{darkpalepink}{\textbf{T6}} Privacy and Security risks} & - & \faCheck & - & - & - & \faCheck & - & \faCheck & \faCheck & \faCheck & \faCheck & \faCheck & - \\
\rowcolor{palepink!10} & \multicolumn{2}{l}{\textcolor{darkpalepink}{\textbf{T7:}} Changes in views, beliefs, morals, and politics} & - & - & - & \faCheck & - & \faCheck & \faCheck & - & - & \faCheck & \faCheck & - & \faCheck \\
\rowcolor{palepink!10} & \multicolumn{2}{l}{\textcolor{darkpalepink}{\textbf{T8:}} Manipulation prevention techniques:} & & & & & & & & & & & & & \\
\rowcolor{palepink!10} &  & \textcolor{darkpalepink}{\textbf{T8.1:}}~\emph{prevent false memories and empathy-based manipulation} & - & - & - & - & - & - & \faCheck & - & - & - & \faCheck & - & - \\
\rowcolor{palepink!10} &  & \textcolor{darkpalepink}{\textbf{T8.2:}}~\emph{security and privacy recommendations} & - & - & - & - & \faCheck & \faCheck & - & - & \faCheck & \faCheck & - & \faCheck & - \\
\rowcolor{palepink!10} &  & \textcolor{darkpalepink}{\textbf{T8.3:}}~\emph{call for research efforts} & - & - & - & - & - & \faCheck & - & - & - & \faCheck & - & - & \faCheck \\
\rowcolor{palepink!10} &  & \textcolor{darkpalepink}{\textbf{T8.4:}}~\emph{improve user literacy} & - & - & - & - & - & \faCheck & - & - & \faCheck & \faCheck & - & - &   \\ \bottomrule
\end{tabular}%
}
\caption{This table presents an overview of our eight themes (e.g.,T2) and 15 sub-themes (e.g., T2.1) and the corresponding publications in which they were mentioned (\faCheck).}
\label{tab:paper-themes}
\end{table}

\subsection{RQ2: How Could XR Amplify the Effects of Deceptive Design?}
\label{subsec:RQ1}

In our analysis, 10 out of 13 publications discussed XR's ability to create an immersive, emotionally engaging, and interactive experience that can be used to amplify user manipulation. 

\subsubsection{T1: Adverse Effects of User Experience}
\label{subsubsec:XR_enhance_User_Experience}

We aggregated four ways XR technologies improve user experience from 10 publications: (1) product previewing, (2) experience previewing, (3) memory remembering, and (4) realistic experience and sensations. However, these publications simultaneously raise concerns about the potential for user deception under the guise of improving user experience. Among these publications, there was a strong agreement that XR technologies can enhance user experience with immersive, interactive, and photorealistic virtual environments. For instance, in their study on XR advertising, ~\citet{mhaidli2021identifying} mentioned the IKEA Place AR-based app~\footnote{\label{note1} See footnote 3.} that lets customers digitally sample furniture in their living area to better comprehend its size, design, and function. Similarly, before booking a hotel room, clients can join a virtual tour to ``sample'' the experience~\cite{mhaidli2021identifying}. These preview features while seemingly convenient, could present an ideal version of reality, potentially leading to uninformed purchases and disappointment when encountering the actual product~\cite{mhaidli2021identifying}. In addition, \citet{bonnail2022exploring}'s hypothesize and \citet{schlembach2021forced}'s analysis of VR documentary suggest that XR has the potential to enhance personal and historical memory reminiscence as it provides ``realistic'' reconstruction of past events in 3D, and~\citet{krauss2022exploring} described a VR documentary (``Meeting You''\footnote{\url{https://welcon.kocca.kr/en/directory/content/meeting-you--4008}}) that enables people to spend time with deceased friends and families ``as if they were still alive.''~\cite[p.~1]{krauss2022exploring}. This also raised ethical concerns regarding the potential for distinguishing between real and false memories, and the potential for impersonation and psychological manipulations~\cite{schlembach2021forced}. Several papers noted that XR technology can produce stereoscopic images, music, and convincing haptic feedback to make people feel psychologically present in a virtual environment~\cite{cummings2016immersive,ramirez2021what,williams2014effects}. Beyond presence, \citet{ramirez2021what} further noted in their theoretical argument paper that carefully designed VR simulations can generate ``virtually real experiences,'' where users engage with the virtual experiences as if they were real. This, achieved through a combination of high context-realism (i.e., the degree to which the rules, settings, and appearance of a simulated world respond to users as if they were real) and perspectival-fidelity (i.e., a simulation that contributes to generating a user's viewpoint), may further blur the line between virual and real and enables XR to be used for manipulation purposes ~\cite{ramirez2021what}.

\subsubsection{T2: XR Exacerbates User Manipulation}
\label{subsubsec:XR_exacertbates_user_manipulation}

Nine publications discussed how XR features increase user manipulation. These publications consistently argue that XR-based deceptive designs are more convincing and manipulative compared to non-XR deceptive designs~\cite{bonnail2022exploring,lebeck2018towards,mhaidli2021identifying,franklin2022virtual}. From our literature analysis, we identified six such XR features that can contribute to user manipulation. 

\paragraph{T2.1: The illusion of objectivity in XR experience.}
To maximize the level of immersion, many XR games and applications promise a first-person experience that allows users ``see'' the virtual world from the perspective of a certain individual, animal, or entity. An example is the \textit{6x9} VR game that simulates a prisoner's first-experience~\cite{Guardian20166x9}. However, \citet{ramirez2021what} argued that XR can never produce an objective first-person experience because the simulations always suffer from \textit{semantic variance}~\cite{selinger2010competence} and \textit{structural intersectionality}~\cite{crenshaw1990mapping}. The potential for creators to inject their own perspectives into the content raises ethical concerns around XR simulations that claim to deliver a first-person experience~\cite{ramirez2021what}. 

\paragraph{T2.2: Obscuring reality, disguising risks.}
As mentioned in~\autoref{sec:background}, XR has the ability to obstruct and substitute the physical objects with a virtual environment. In a multi-user scenario, \citet{buck2022security} discussed the potential for one user to mislead another regarding their physical surroundings. This concern was echoed by \citet{lebeck2018towards} through AR user interviews, which illustrated users' concern that XR allows hiding of dangerous physical items (e.g., a gun) behind virtual objects, causing harms to people using XR. However,~\citet{mhaidli2021identifying} noted that in XR, users cannot just look away like on a 2D screen, especially when they are unaware of the threats. 

\paragraph{T2.3: Data fuels privacy risks and manipulation.}
Several publications emphasized that XR technologies process lots of data to support its user experience, such as data describing user body and eye movements~\cite{buck2022security,lebeck2018towards,lee2021adcube}, gestures~\cite{lee2021adcube,mhaidli2021identifying}, physical appearance~\cite{mhaidli2021identifying,buck2022security}, behavioural patterns (e.g., gait, mannerisms)~\cite{mhaidli2021identifying,lebeck2018towards,buck2022security}, and physical surroundings~\cite{lebeck2018towards, buck2022security,mhaidli2021identifying}. The user experiments conducted by~\citet{miller2020personal} and ~\citet{pfeuffer2019behavioural} revealed that XR users can be deanonymized based on their behavioral patterns, a type of data that commonly considered non-private. These data can be a valuable resource for predicting user preferences and vulnerabilities, which enables the development of tailored manipulation strategies~\cite{mhaidli2021identifying}. 

\paragraph{T2.4: Unawared access, unawared data use.}
Beyond single-user environments, three studies focused on multi-user interaction and application data misuse. For instance, the user studies conducted by~\citet{lebeck2018towards} documented cases where users modify each others' choice architectures by manipulating (e.g., hiding) each other's virtual objects. Moreover, \citet{lebeck2018towards} revealed several problematic multi-user behaviours, such as when users place inappropriate virtual objects in the shared space or onto others' avatars or damage others' virtual objects without permission. Regarding information misuse, \citet{lebeck2018towards} identified users' concerns regarding accidental disclosure of private information without understanding others can see it and bystanders' concerns of being unwilling participants in others' XR experiences. \citet{buck2022security} further warned that these data might be abused by the XR applications or users, causing unexpected privacy risks and social consequences. This theme differs from the previous one in that the user was neither actively involved nor aware of the data collection process.

\paragraph{T2.5: Insecurity worsens user manipulation and privacy concerns.}

Several papers highlighted that current XR systems lack security controls that usually exist on mobile, PC or web environments, leaving users vulnerable to manipulation and data privacy threats. For instance, due to the absence of the Same Original Policy (SOP) in WebVR~\cite{Mozilla2020SOP}, ~\citet{lee2021adcube} prototyped a defense system to prevent attackers to inject ads or alter content on VR webpages~\cite{lee2021adcube}. Moreover,~\citet{su2022perception}'s experiment demonstrated cases where attackers could steal user information by side-channelling the VR hardware, and \citet{buck2022security} expressed concerns about the migration of Internet scams into the metaverse. 

\paragraph{T2.6: Persistent exposure to manipulation.}

Two papers further expressed the concerns regarding user manipulation from a constant immersive experience~\cite{mhaidli2021identifying,cummings2022all}. As companies are now developing daily-use XR equipment (i.e., Mojo smart lens~\footnote{Mojo smart contact lens that overlays virtual information onto the physical world. \url{https://www.mojo.vision/mojo-lens}}), ~\citet{mhaidli2021identifying} hypothesized that future users will worn AR devices constantly throughout their daily life. As a result,~\citet{cummings2022all} and~\citet{mhaidli2021identifying} warned that users will experience constant data collection and user surveillance, and their exposure to targeted advertising and XR-based manipulation will also become persistent.

We argue that Theme 2 revealed XR's potential to enable deceptive design across all six attributes~\cite{mathur2021makes,mathur2019dark}. The illusion of objectivity can be employed to induce false beliefs through users' perceived objectivity of their experiences. XR's capacity to obscure reality may facilitate ``information hiding.'' User data from XR sensors enables ``covert'' and ``asymmetric'' deceptive design that exploits users' cognitive biases and preferences. The insecurity of XR systems allows for all types of deceptive design that exploits XR design elements and users' false beliefs in the authenticity of XR content.

\subsection{RQ3: What Deceptive Design Strategies Could Presented in XR?}
\label{subsec:RQ2}

Ten publications detailed XR deceptive design strategies, but none offered a taxonomy. As presented in \autoref{tab:paper-themes}, we classified those deceptive design strategies into three themes: (1) Psychological manipulation, (2) Reality distortion, and (3) Tricking user perception. In this section, we introduce each theme, and how the manipulation is enhanced by the XR features.

\subsubsection{T3: Psychological Manipulation}
\label{subsubsec:RQ3_Theme1_Psy}

Eight publications discussed psychological manipulation through XR technologies~\cite{mhaidli2021identifying,schlembach2021forced,bonnail2022exploring}, which we detail in this section.

\paragraph{T3.1: False memory implantation.} 

A central feature of XR technology is \textit{immersion}, where users are fully engaged and absorbed within a virtual environment~\cite{XRSI2020Definition}. \citet{bonnail2022exploring} and \citet{schlembach2021forced} emphasized the memory manipulation risks posed by the immersive experience. As detailed in Theme 1 (\autoref{subsubsec:XR_enhance_User_Experience}), XR is capable of supporting memory reminiscing with ``realistic'' multi-dimensional reconstructions. However,~\citet{bonnail2022exploring} described the theory behind human memory flaw and argue that distorted XR reconstructions can lead to false memories. Specifically, humans suffer from \textit{source confusion} and thus may confuse where their memories~ originated~\cite{shapiro1991making}. For example, a conversational rumour may be mistaken for TV news~\cite{bonnail2022exploring,shapiro1991making}. Based on a UK VR advertising company's ``memory relive'' service (Momento\footnote{Memento is a UK advertising company that uses AI and VR to enable people to capture and relive their memories.~\url{https://www.mementovr.com}}),~\citet{bonnail2022exploring} hypothesized a scenario of a VR wedding reconstruction, where the VR service provider altered the champagne and wedding dress brands and customers falsely believed they enjoyed this champagne during their wedding and bought from the brand again.

\paragraph{T3.2: Artificial prosthetic memory and empathy-based manipulation.}
Drawing on the literature on memory,~\citet{schlembach2021forced} mentioned that VR can be used to create \textit{prosthetic memories} of historical events. \textit{Prosthetic memory} is a public cultural memory people acquire from movies and television~\cite{landsberg2004prosthetic}. It fosters a personal connection to others' historical experiences, and can create people's empathy, social responsibility, and political alliances across race, class, and gender~\cite{landsberg2004prosthetic}. For example, people who had no personal connection to the 9/11 victims can became vicariously traumatized after watching broadcast media and films that exposed them to victims' sufferings~\cite{kaplan2005trauma}. Humanities and social sciences literature believed that VR offers a route to empathy because it allows people to ``walk in another’s shoes'' and ``see the world through another’s eyes.''~\cite{rose2018immersive} Through the examination of two VR films, \citet{schlembach2021forced} warned against ``empathy generation'' through VR technology as people can be ``made to `feel' something they will be changed by it and so will their behaviour.'' They illustrated this argument through the example of a VR game (\textit{6x9}) that simulates the psychological effects of extreme isolation resulting from solitary confinement, including distorted vision, hallucinations, floating, and screaming~\cite{Guardian20166x9,schlembach2021forced}. This immersive experience evoked public empathy towards prisoners and led to the reform of legal and political decisions (i.e., in 2016, the Obama administration introduced federal changes to curb solitary confinement)~\cite{USDJA2016Obama}. 

In parallel, drawing from literature on the ethics of choice-architecture, \citet{ramirez2021what} argued that the VR-based empathy enhancing simulations are always unethical because it can never objectively provide actual experience of another, as we discussed in Theme 2 in~\autoref{subsubsec:XR_exacertbates_user_manipulation}. Thus, it may mislead users ``about the nature of their virtual experiences,'' and manipulate them into ``changing their thoughts, beliefs, and behaviours,'' consequentially influencing their ``future judgments about their own values, policies, and so on''~\cite[p.~530--532]{ramirez2021what}. In addition, taking inspiration from a U.S. military recruitment game (i.e., America's Army~\footnote{America's Army is the official game of the U.S. Army.~\url{https://dacowits.defense.gov/Portals/48/Documents/General\%20Documents/RFI\%20Docs/Dec2018/USA\%20RFI\%203\%20Attachment.pdf?ver=2018-12-08-000554-463}}), \citet{mhaidli2021identifying} hypothesized that such games in VR may be more powerful in driving players' inclination towards enlist. They argued that immersive and realistic experiences of VR could create a sense of "living" in the military, and the enjoyment of the game might ``bias'' players to view joining the armed forces positively~\cite{mhaidli2021identifying}.

\paragraph{T3.3: Hyper-personalized user manipulation.}
\textit{Personalization} is an e-commerce strategy that employs data to enhance experience of people with similar traits through tailored content, services, and features~\cite{mhaidli2021identifying,vesanen2007personalization}. \textit{Hyper-personalization} refers to creating highly customized content tailored to an individual~\cite{The2022Williams, mhaidli2021identifying}. As detailed in Theme 2 (\autoref{subsubsec:XR_exacertbates_user_manipulation}), XR's data collection capabilities enable inference of user states and facilitate manipulation that takes advantage of user data. Drawing inspiration from IKEA's AR furniture preview application\footnote{See footnote 3.},~\citet{mhaidli2021identifying} discussed how AR-driven hyper-personalization can manipulate people's purchase decisions by making ``the photorealistic rendering of the furniture seem brighter and more colorful than real life.'' Two papers echoed this discussion and illustrated other possible cases of data-based user manipulation in XR~\cite{buck2022security,mhaidli2021identifying}. For example, psychological user manipulation through idealized interaction partners (meeting their preferences) based on biometric data~\cite{buck2022security}, emotional manipulation in user investment decisions using holographic recreations of their trusted people based on shared images~\cite{mhaidli2021identifying}, and physical navigate the user to a restaurant based on physiological signal of hunger~\cite{mhaidli2021identifying}.

In summary, we classify this Theme as ``deceptive,'' ``covert,'' and ``information hiding''~\cite{mathur2021makes,mathur2019dark} as it takes advantage of users' false belief in the authenticity of XR simulations, hides information about the ads inclusion, and manipulates them through mechanisms such as the photorealistic graphics and sensory feedback that are not directly apparent. We also argue that this theme partially but not entirely align with the pre-established ``asymmetric'' attribute in~\citet{mathur2021makes} as the XR simulation subtly influence user decisions without any interface friction.

\subsubsection{T4: Reality Distortion}
\label{subsubsec:RQ3_Theme2}

Several publications discussed user manipulation through seamless reality distortion and the idea that XR equipment will be worn daily~\cite{mhaidli2021identifying,bonnail2022exploring,lebeck2018towards}. Drawing upon the deceptive ads in the 2020 U.S. presidential campaign,\footnote{During the 2020 U.S. presidential campaign, both candidates issued misleading ads featuring deceptive statements and distorted images.~\url{https://cnn.com/2021/11/08/politics/fact-check-house-republican-ad-trump-images-2020/index.html}}~\citet{mhaidli2021identifying} suggested a potential scenario where future politicians could exploit daily-worn XR devices, concealing evidence of poverty and economic downturn in a city with photorealistic graphics. ~\citet{bonnail2022exploring} echoed this scenario by proposing a potential use of VR technologies for marketing, where people's uploaded photos could be modified to incorporate brand logos and create false memories of experiencing a product. ~\citet{lebeck2018towards}'s AR user study further emphasized users' safety concerns in AR-assist driving, where deceptive representations of reality, such as hidden pedestrians or misrepresented neighbouring vehicle, could cause harms to both the driver and people nearby. The distorted reality can lead to psychological addiction because people ``identify with the bodies that they virtually inhabit, even when those bodies are very different from their own''~\cite[p.~532]{ramirez2021what}.

We classify this theme as ``asymmetric,'' ``covert,'' and ``deceptive'' because it increases people's exposure to ads, utilizes subtle influence mechanisms in XR that are unapparent to users, and capitalizes on people's false beliefs by modifying the representation of reality~\cite{mathur2021makes,mathur2019dark}.

\subsubsection{T5: Tricking User Perception}
\label{subsubsec:RQ3_Theme3}

While our Theme 3 (\autoref{subsubsec:XR_enhance_User_Experience}) discussed XR's ability to offer immersive virtual experiences, four publications highlighted its manipulation on user perceptions~\cite{lee2021adcube,su2022perception,buck2022security,mhaidli2021identifying}. 

\paragraph{T5.1: Extreme Realism blurs the boundary between virtual and reality}

\citet{mhaidli2021identifying} suggested that a skillfully crafted commercial with highly realistic visuals might fool viewers into thinking it was real. AR users may not realize that other people's shirts include digital brand logos. Although current XR systems cannot yet generate such realistic simulations, \citet{mhaidli2021identifying} and \citet{buck2022security} anticipated that it will be possible in the near future. In addition, \citet{mhaidli2021identifying} and \citet{krauss2022exploring} noted XR's product previewing feature to sway user buying decisions (e.g., brighter and more colourful furniture). The virtually real experience in XR makes the preview seem authentic, resulting in disappointment when the product is lower quality ~\cite{mhaidli2021identifying}.

\paragraph{T5.2: Perception hacking manipulates user input}

Using XR technologies' visual, auditory, and haptic feedback to engage with the virtual world, \citet{su2022perception}' lab experiment and \citet{lee2021adcube}' user study both showed that \textit{perception hacking} in XR can effectively deceive people. An well-known non-XR example of perception hacking is \textit{Clickjacking}, which misleads users into accidental pressing of ads on websites (i.e., disguised ads)~\cite{Brignull2022deceptive,clickjacking2022Kantor}. Our analysis synthesized three types of XR perception hacking: (1) cursor-jacking~\cite{su2022perception,lee2021adcube}, (2) blind spot tracking~\cite{lee2021adcube}, and (3) auxiliary display abuse~\cite{lee2021adcube}.

As a variation of clickjacking on the web, \textit{Cursor-jacking} in XR manipulates visual cues of object movements (e.g., hand movements) to differ from proprioceptive cues (e.g., physical sensing of hand movements)~\cite{su2022perception,lee2021adcube}. Research indicated that visual perception is valued more than other senses, even if the difference is small~\cite{kohli2012redirected, azmandian2016haptic}. ~\citet{lee2021adcube}'s user study showed that an angular offset of visual cues from the user's hands can redirect hand movement in VR and deceive them into interacting with undesired virtual objects. Taking inspiration from ad impression frauds on websites and mobile apps~\cite{stone2011understanding,liu2014decaf}, in a 360-degree immersive XR environment, \citet{lee2021adcube} demonstrated \textit{blind spot tracking} that uses gaze analysis to display ads beyond people's line of sight. Similar to blind spot tracking,~\citet{lee2021adcube} also demonstrated \textit{auxiliary display abuse} that occurs when participants immersed in XR and therefore fail to notice ads on a secondary monitor (e.g., PC monitor)~\cite{lee2021adcube}.  

In short, we classify this Theme as ``asymmetric,'' ``covert,'' ``deceptive,'' and ``information hiding''~\cite{mathur2021makes,mathur2019dark} as it increases users' exposure to specific brand logos, trick users into interacting with undesired virtual objects through subtle influence mechanisms, and capitalizes on people's false beliefs of perceived body movements and hide the existence of ads outside of their line of sight.

\subsection{RQ4: What Risks Could XR Deceptive Design Pose to Users?}
\label{subsec:RQ3}

We synthesized two risks XR deceptive design poses to users: (1) privacy and security risks, and (2) change to users' feelings, behaviours, and moral and political orientations. This section discusses both risks and how they are unique to XR. Eight publications proposed solutions to prevent XR deceptive design~\cite{franklin2022virtual,nijholt2012trends,lebeck2018towards,cummings2022all}. We synthesized (3) XR manipulation prevention strategies, and discussed them at the end of this section. 

\subsubsection{T6: XR deceptive design worsens privacy and security risks.}

Several publications expressed a concern that deceptive designs in XR may lead to greater and harder-to-defend privacy loss compared to those in non-XR~\cite{bonnail2022exploring,cummings2022all,krauss2022exploring}. For example, to enjoy the XR memory reconstruction service, customers must share personal images and video recordings, as ``the more video provided, the more accurate the construction''~\cite[p.~3]{bonnail2022exploring}. Based on diverse user data, XR produces superior experiences compared to non-XR systems, making it more challenging for users to prioritize privacy~\cite{cummings2022all,sun2021calculus}. In addition, XR applications can deceive users into disclosing private information to digital humans or those who resemble their loved ones~\cite{buck2022security,mhaidli2021identifying,krauss2022exploring}. Further,~\citet{cummings2022all} proposed a possibility where privacy concerns may hinder users from reporting encountered deceptive design, potentially providing a ``shield'' for such issues in XR.

There was also a strong agreement that deceptive design in XR can lead to security risks~\cite{su2022perception,buck2022security,lebeck2018towards}. For example, \citet{su2022perception} and \citet{buck2022security} proposed that perception hacking (Theme 5 in~\autoref{subsubsec:RQ3_Theme3}) can deceive people into accidental granting of privileges. \citet{lee2021adcube} further noted that the insecurity of XR browsers leave space for ad fraud, and \citet{su2022perception} mentioned that attackers can exploit these security weaknesses to steal sensitive user data or distort XR content. 

\subsubsection{T7: XR deceptive design leads to changes in users' views, beliefs, morals, and politics.}

Another theme frequently mentioned by the publications is that misleading design induces artificial feelings that change people's judgement of values, views, beliefs, morals, and politics~\cite{mhaidli2021identifying,schlembach2021forced,franklin2022virtual}. For instance, ~\citet{schlembach2021forced}'s analysis of the \textit{6x9} VR simulation game~\cite{Guardian20166x9} elicited that empathy-based manipulations are effective in political and social control. ~\citet{mhaidli2021identifying} mentioned another example of concealing poverty and economic decline to show a false bright economic environment for political campaigns. Users' false beliefs in XR's objectivity lead to ``made'' empathetic responses to manipulative simulations, which in turn, affects their future thoughts, attitudes, and emotions~\cite{ramirez2021what}. Furthermore, the idealized digital humans, as suggested by~\citet{buck2022security} and~\citet{franklin2022virtual}, can potentially be used for ``political duress'' or manipulating users into engaging in risky or unlawful behaviours, such as online gambling.

Several publications mentioned that people's manipulated attitude, belief, thought, and perception can influence their behaviours~\cite{ramirez2021what, lebeck2018towards, lee2021adcube, buck2022security,franklin2022virtual}. For instance, the distorted preview of products and experiences~\cite{mhaidli2021identifying}, the use of hyper-personalized ad spokesman~\cite{mhaidli2021identifying, buck2022security}, and the integration of ads in XR memory reconstructions can impact people's buying decisions~\cite{bonnail2022exploring}.  In addition, ~\citet{franklin2022virtual} gave an example that players' behaviours in video gameplay can ``spillover'' into their behaviours outside of game (i.e.,\textit{Virtual spillover}~\cite{quwaider2019impact}). Thus, they argued that the same effect may occur among XR users, especially when XR exposure becomes constant~\cite{franklin2022virtual}.

\subsubsection{T8: XR manipulation prevention techniques}
\label{subsubsec:RQ4-theme3}

Our analysis found eight publications suggesting solutions to protect people from XR deceptive design~\cite{nijholt2012trends,lebeck2018towards,bonnail2022exploring,ramirez2021what}. For example,~\citet{franklin2022virtual} and~\citet{nijholt2012trends} recommended using speech and text analysis, physiological sensors, and intelligent software to detect and remove behavior- and preference-changing methods. To address the multi-user problems,~\citet{lebeck2018towards} suggested creating methods and policies to prevent XR devices from overlaying virtual on others' content and allowing users control over view and edit permissions for personal virtual objects and space. 

\paragraph{T8.1: Preventing false memories and empathy-based manipulation.}
In the cases of memory manipulation, \citet{bonnail2022exploring} proposed basing XR memory reconstruction on user consent, with service providers openly disclosing potential modifications, and identifying updated items in scenes to prevent source confusion and false recollections. Since empathy-based manipulation plays a crucial role in XR deceptive design, \citet{ramirez2021what} proposed changing simulation goals from empathy-provoking first-person experiences to bystander perspectives. This reduces manipulation and engages users sympathetically instead of empathetically~\cite{ramirez2021what}.      

\paragraph{T8.2: Security and privacy recommendations.}
Five publications provided privacy and security recommendations for developing future XR technologies~\cite{cummings2022all,buck2022security,lee2021adcube}. For example,~\cite{cummings2022all} indicated that XR users needs to trust the information they consume and the services that use their data to freely and safely share information. ~\cite{buck2022security} argued that power should be given to the user, and they should be able to opt-out of data collection at any moment. In addition,~\cite{su2022perception} advocated for enhancing security in XR systems, and ~\cite{lee2021adcube} called for creating XR-specific security policies. ~\cite{cummings2022all} specifically mentioned that children should be protected with special laws and standards, and three publications agreed that XR data collection and misuse for malicious purposes should be prohibited~\cite{buck2022security,cummings2022all,mhaidli2021identifying}. 

\paragraph{T8.3: Call for academic research efforts to better understand the problems.}
Three studies called for research about problems in XR deceptive design~\cite{buck2022security,franklin2022virtual,mhaidli2021identifying}. For instance, \citet{buck2022security} suggested researching on psychological user manipulation in XR ads and the effects of physical properties of avatars. \citet{buck2022security} and \citet{franklin2022virtual} recommended researching how digital interaction translates into people's daily lives. \citet{mhaidli2021identifying} called for research on what preference changes are ethical or manipulative based on user benefits, and proposed studying privacy policies of XR advertisers and developers to understand XR data risks. 

\paragraph{T8.4: Improve user literacy to identify and resist manipulation}
Three publications emphasized educating users to raise awareness and resilience towards deceptive design and its risks~\cite{mhaidli2021identifying,buck2022security,cummings2022all}. Specifically,~\citet{mhaidli2021identifying} proposed improving users' literacy of recognizing photorealistic XR ads with distorted previews and hyper-personalized avatars. They argued that enhancing users' literacy in XR ads could improve their resistance to manipulation~\cite{mhaidli2021identifying}. \citet{buck2022security} and \citet{cummings2022all} called for user education on XR data flow and privacy implications. According to \citet{mhaidli2021identifying}, developing effective educational interventions will require academic and industrial research.

\section{Discussion and Takeaways}
\label{sec:discussion}
Our systematic analysis of the literature synthesized existing research in XR deceptive design. We identified deceptive design definitions in XR studies, XR features that can amplify user manipulation, XR design patterns demonstrated deceptive attributes~\cite{mathur2019dark,mathur2021makes}, potential risks to users, and solutions proposed in the existing literature. In this section, we outline directions for future research and offer recommendations to XR designers and policymakers.

\subsection{Comparing Deceptive Design in XR and Other Platforms}
 
Our eight themes of XR deceptive design have many convergence with established taxonomies in deceptive design research on websites~\cite{mathur2021makes,gray2018dark,gray2023towards}, games~\cite{zagal2013dark,mathur2021makes}, and 3D interactions~\cite{Greenberg2014proxemic}. For instance, we found possible XR variations of ``toying with emotion''~\cite{gray2018dark}, such as artificial prosthetic memory that exploits empathy~\cite{schlembach2021forced} and hyper-personalized interaction partners that evoke positive feelings~\cite{mhaidli2021identifying}. The latter may also be viewed as a form of ``social engineering''~\cite{gray2023towards} and ``attention grabber'' because people tend to focus on content aligned with their interests~\cite{Greenberg2014proxemic}. In addition, digital humans mimicking people's trusted and loved ones can be viewed as a form of ``friend spam'' in the XR~\cite{zagal2013dark,gray2018dark}. The risks that XR data may be used for user profiling is a form of ``disguised data collection''~\cite{Greenberg2014proxemic}. While not fully aligning with established taxonomies, the perception hacking in XR resembles interface strategies such as ``sneaking'' or ``forced action'' that deceive users into engaging with unwanted virtual objects and behaviours~\cite{gray2023towards,gray2018dark}. On the other hand, our analysis also unveiled possible new forms of deceptive design emerged from XR's immersive nature. Examples include false personal memory implantation, blind-spot tracking, and reality distortion that capitalizes on human memory flaw and visual limitation, and the photorealistic simulation, 360-degree display, and the coexistence of virtual and reality in certain XR systems. 

Using \citet{mathur2021makes}'s attribute classification, our literature analysis showed that the majority of deceptive design in XR demonstrated attributes such as ``covert,'' ``deceptive,'' and ``information hiding,'' similar to the major attributes demonstrated by web-based deceptive design~\cite{mathur2019dark}. However, in XR, deceptive design employs subtler influence mechanisms and can more accurately target people's weaknesses based on extensive data. The resulted manipulations are novel, unfamiliar, and challenging for people to detect.

\subsection{Future Research Directions in XR Deceptive Design}

Research in deceptive design is maturing, and XR has experienced rapid growth in the past few years. However, in-depth investigations of deception mechanisms that leverage XR's immersive capabilities require more study. In the following, we highlight the most urgent action items that need research attention. We hope that our results provide a direction for future research to identify new manipulations in XR and develop appropriate countermeasures. 

\subsubsection{A better understanding of deceptive design in XR}

The findings of RQ1 (see \autoref{subsec:overview}) showed that existing research primarily relied on deceptive design definitions that emphasize designer intentions (e.g.,~\cite{nijholt2012trends,buck2022security,torstensson2020wizard}). This trend shows a need for XR deceptive design research using an updated definition and focusing on those that resulted from unintentional design decisions. Moreover, to effectively mitigate harmful deceptive design in XR, it is critical to understand to what extent the design mechanisms are considered benevolent (e.g., nudging) and to what extent they are considered deceptive (e.g., tricking). \citet{gray2021enduser} and \citet{gray2023mapping} suggested that users' perceptions of manipulation can help identify borderline practices that are at may not be strictly problematic or illegal. During our literature analysis, we found papers that tried to differentiate ``nudging'' and ``manipuation'' based on user interest and freedom of choice (e.g.,~\cite{mhaidli2021identifying, ramirez2021what}). Therefore, we suggest that future deceptive design research should include a comprehensive user-centric interaction or perception analysis. This may include exploring questions such as: How do users perceive XR deceptive design for different purposes? Which designs are considered manipulative and need to not be built, and which are acceptable? Without a clear understanding of users' perspectives, developing effective and user-friendly countermeasures to deception will be challenging.      

Our analysis uncovered deceptive designs from non-XR environments that are now extended to XR. We also observed new forms of deceptive designs emerged from XR features. However---given that XR is still in its infancy and XR devices have yet to become mainstream---most of these studies were based on hypothetical scenarios of potential future problems. More problematic issues are likely to emerge with the advancement of XR technology. Therefore, we call for further research on new forms of XR deceptive designs and their impacts on users. That way, countermeasures can be developed to address these challenges before they can seriously harm people.     

\subsubsection{A more transparent awareness of XR data use practices}

Among the publications we reviewed in our study, there was a noteworthy agreement that XR data enables deceptive design~\cite{buck2022security,lebeck2018towards,miller2020personal}. The extensive collection of user data through XR devices creates opportunities for precise user profiling, which can be used for hyperpersonalization and manipulation~\cite{mhaidli2021identifying,buck2022security,su2022perception}. To better comprehend and prevent this issue, we suggest that future research should start by asking what data types can be collected by XR sensors. The investigation by the XR Safety Initiative (XRSI) research lab's \textit{1st XR data classification roundtable} conference~\cite{XRSI2020classification} and \citet{dick2021balancing} in 2021 have laid the groundwork for examining how VR and AR use data. However, several questions still need to be answered. These include questions such as: What user information can be inferred from the XR sensor data (especially from MR sensors)? How can these data be exploited in XR deceptive design? How does the use of these data impact user autonomy, safety, and privacy? What controls do users want to have over these data? Addressing these questions and developing appropriate protections for XR technology users require the continued efforts of researchers, developers, and designers.

\subsubsection{A deeper education on deceptive design practices in XR}

Lastly, our findings in T8 (\autoref{subsubsec:RQ4-theme3}) emphasized the need for user education on deceptive design in XR~\cite{mhaidli2021identifying,buck2022security,cummings2022all}. Research has shown that users, despite being able to recognize deceptive design, often have a resigned attitude due to their dependence on the services~\cite{maier2019dark}. Conversely, ~\citet{bongard2021manipulated} found that users' vague awareness of the entailed concrete harms on themselves made them failed to realize the necessity of taking self-protective actions. Therefore, we argue that it is crucial to design effective user educational interventions. Such interventions should instruct people that XR may not always reflect reality. Users should be taught about cues that differentiate fictional simulations from realistic representations. Moreover, it is essential to inform people about XR data tracking, potential privacy issues, and XR mechanisms that could manipulate them to share their data. Most importantly, users should understand the possible consequences and risks of XR deceptive design. They should also receive training to help them detect scams, digital (i.e., fake) humans, and bots that pretend to be someone they trust. Future research is also needed to explore optimal educational content design for improving user understanding, measuring user learning outcomes, and assessing user resilience against deceptive design in XR.

\subsection{Implications to Designers}

Previous studies have given several design recommendations for how to mitigate empathy-based deceptive design in immersive environments. As discussed in \autoref{subsubsec:RQ4-theme3}, \citet{ramirez2021what} proposed promoting sympathy-based feelings instead of empathetic feelings when designing immersive environments. Other scholars have recommended implementing auditing systems to detect manipulative designs~\cite{franklin2022virtual,nijholt2012trends}. We believe that all these suggestions require more research to fully understand how deceptive design is evolving in XR and how design mechanisms affect users' perceptions of empathy and sympathy.

Additionally, \citet{cummings2022all} and \citet{mhaidli2021identifying} both highlighted the critical role of trust and transparency in XR content. With this in mind, we think that future XR design should make it easier for people to distinguish the difference between fiction simulations and reality representations. One approach could be to incorporate visual cues within the simulations~\cite{bonnail2022exploring}. This technique prevents people from getting confused and ensures they can distinguish fact from fiction. 

Several studies have emphasized how essential user empowerment is in XR systems. This includes integrating user consent, giving users control over their personal virtual objects and space, and transparently informing and enabling users to opt-out of data collection through XR devices. Previous research has shown that users are not always against collecting data, but rather against collecting sensitive or non-consensual data~\cite{naeini2017privacy}. As the number of XR sensors continues to grow, we further suggest that XR systems should be set up for people to opt-in, rather than opt-out of sharing information. Systems should also provide people with clear instructions on how to revoke consent, access to, and control over the collected data.

Although XR privacy concerns have received attention in the literature, many articles have also highlighted how crucial it is to implement adequate security controls to protect users from potential cyberattacks and manipulations. This can be achieved through integrating established security measures found in non-XR technologies~\cite{lee2021adcube, su2022perception}. Thus, we further proposed that specific protections should be developed for different user groups. For instance, because XR is being used more in high-school education, it is important to implement child-specific protections.

\subsection{Implications to Policymakers}

We found that the results of several publications have implications for creating security and privacy policies for XR. These implications go beyond design recommendations.

Our findings in \autoref{subsubsec:RQ4-theme3} revealed a need for regulating the collection and use of XR data, especially data from unwillingly involved parties (e.g., other users and bystanders). A recent analysis of privacy regulations shows that current frameworks are inadequate in addressing the possible risks of XR technologies~\cite{dick2021balancing}. The challenge of balancing innovation with regulation is not a novel issue. Technology frequently outpaces regulatory development. In light of the privacy concerns XR technologies raise, we thus suggest policymakers conduct a thorough review of how well existing policies and regulatory frameworks work in the XR space. The assessment of US regulations~\cite{dick2021balancing}, XR user privacy concern and protection-seeking strategies~\cite{hadan2024privacy}, and the Oculus VR privacy policies~\cite{trimananda2021ovrseen} serve as a starting point. However, there needs to be more research into regulatory frameworks in other countries and the policies for other XR devices and applications. We suggest that policymakers consider questions like whether existing rules and policies can protect people's privacy and security, especially for children, within the XR domain. 

Creating an immersive experience in XR requires collecting user data~\cite{buck2022security,cummings2016immersive}. From the publications we analyzed, we have identified the need for XR-specific security policy~\cite{lee2021adcube}, and how regulating malicious data use is necessary~\cite{buck2022security, cummings2022all}. An analysis found 70\% of Oculus VR data flows were not properly disclosed and 69\% were used for purposes unrelated to the core functionality of the device~\cite{trimananda2021ovrseen}. Thus, we suggest that policymakers should develop regulations that foster the development of these technologies while preventing data use for manipulation, and should focus on reinforcing existing regulations and ensuring that companies comply with what they promised.

Lastly---as a part of user empowerment---we propose that XR policies and consents that are presented to users should be easy-to-read (e.g., as required by the California Consumer Privacy Act (CCPA)\footnote{California Consumer Privacy Act (CCPA).~\url{https://oag.ca.gov/privacy/ccpa\#:~:text=The\%20CCPA\%20requires\%20businesses\%20to,use\%20the\%20categories\%20of\%20information.}} for non-XR technologies), so that the people can quickly understand the information without having to learn complicated terms.

\subsection{Limitations}
Our research is limited in multiple aspects. The nascent nature of deceptive design research in XR has resulted in a scarcity of existing literature in this field. Consequently, our literature review is based on limited resources. In addition, given the novelty of XR technology, most of these publications only discussed \textit{hypothetical} scenarios derived from non-XR use cases. Nonetheless, we consider our preliminary review of XR deceptive design literature to be crucial for several reasons. First, the anticipated rapid development of XR technology suggests that the potential deceptive design scenarios and their associated concerns identified in this study are likely to materialize in the near future. Second, the limited sample size of our literature review highlights the critical need for further academic inquiry into the complexities and potential harms of deceptive design within XR environments. By fostering a deeper understanding of these issues, we can pave the way for the development of effective strategies to safeguard users from potential exploitation as malicious practices within XR domains begin to take root.

Regarding our methodology, we acknowledge the previously reported instability in literature database search results, as evidenced by previous works (e.g.,~\cite{rogers2022much,macarthur2021you}), where searches conducted on different days resulted in different numbers of records. To address this issue, we have adjusted the search syntax for each of the four literature databases selected. We also used a combination of search queries and user interface features to maintain consistency across all databases. Our search was conducted on multiple days to guarantee the accuracy and comprehensiveness of the results. Despite these efforts, we acknowledge that our outcomes might not fully eliminate this limitation's possible impact.   

We also note that three coders conducted our thematic synthesis, and thus, the resulting themes were shaped by their perspectives and experience in HCI, games, and privacy and cybersecurity research. While we acknowledge the potential for bias, we also recognize the inherent strength this diverse background brings to the analysis. The coders' experiences equip them with valuable insights and frameworks for interpreting the core issues associated with XR deceptive design. Nevertheless, this potential for bias should be considered when interpreting the study's findings.

Finally, we acknowledge that our research focuses on the early stages of XR deceptive design. It aims to create a foundational understanding and set the stage for future investigations. We recognize that as the field matures and further scholarship emerges, a re-examination of the literature may reveal a more comprehensive and holistic perspective on XR deceptive design knowledge.

\section{Conclusion}
\label{sec:conclusion}

In this paper, we presented the results from a systematic analysis of XR deceptive design research. Following the PRISMA guideline~\cite{page2021prisma}, we started with 187 publications from four bibliography databases and analyzed 13 relevant papers published between 2012 to 2022. Our analysis identified eight themes that answer four RQs. The first RQ identified how XR researchers define and distinguish deception from related concepts such as nudging and persuasion. This highlights the need for future researchers to focus more on XR deceptive design unintentional design decisions and the difference between manipulative and benevolent design strategies. The second and the third RQs, answered by Themes 1-5, explored the novel forms of deceptive design potentially emerging from XR's immersive nature, unique interface elements, and sensory feedback. For example, we found potential for creators to exploit users' sense of objectivity by injecting their perspectives into the XR experience. In addition, the vast amount of user data collected by XR technologies could allow for hyper-personalized manipulation strategies, targeting people's vulnerabilities with high accuracy. Furthermore, the ability of certain XR technologies to block out reality raises concerns about potential reality distortion and user perception manipulation. The fourth RQ, answered by Themes 6-8, synthesized the potential impacts of deceptive design in XR on users and explored potential prevention techniques proposed in the literature.

Building upon the synthesized knowledge, we propose actionable recommendations for future research, policymakers, and XR designers to address the challenges posed by deceptive design in XR environments. Our analysis revealed both convergence and divergence between existing XR deceptive design research and established taxonomies from other platforms. Notably, XR facilitates the deployment of novel and subtle deceptive design strategies that exploit the XR's unique features and more precisely target user vulnerabilities. The potential for deceptive design using XR data highlights the need for further research to understand and regulate XR data practices and mitigate associated user risks. It is crucial to acknowledge that this study focuses on the early stages of XR deceptive design research. As the field matures, expanded examinations of the literature may reveal additional insights and complexities. 

Our findings further translate into actionable implications for XR designers and policymakers. For designers, we advocate for fostering sympathy-based responses over empathy-based manipulation,  and the implementation of auditing systems to identify potentially deceptive design elements is crucial. Building trustworthy and secure XR environments alongside empowering users with transparent data practices through XR design are also important. For policymakers, our findings highlight the need for regulations governing the collection and use of XR data. It is also critical to establish regulatory frameworks that adequately address potential risks associated with XR technologies, while avoiding hindering its innovation. We hope that by providing insights from existing research, this paper can serve as a foundational guide for researchers, designers, and policymakers. This will lead to the development of XR experiences that harness technological potential while prioritizing user safety, privacy, and ethical considerations.

\begin{acks}
We would like to thank graduate researchers Sabrina Alicia Sgandurra and Derrick Wang for their insightful feedback on the manuscript and their expertise in resolving technical issues during our paper formatting. We also thank the reviewers for their constructive criticism that ensured a smooth and polished final publication of our research. The research was supported by the Natural Sciences and Engineering Research Council of Canada (NSERC) Discovery Grant (\#RGPIN-2022-03353 and \#RGPIN-2023-03705), the Social Sciences and Humanities Research Council of Canada (SSHRC) Insight Grant (\#435-2022-0476), the Canada Foundation for Innovation (CFI) JELF Grant (\#41844). Any opinions, findings, and conclusions or recommendations expressed in this material are those of the author(s) and do not necessarily reflect the views of the NSERC, SSHRC, the CFI, nor the University of Waterloo.
\end{acks}

\bibliographystyle{ACM-Reference-Format}
\bibliography{References}


\end{document}